\documentstyle[prd,aps,preprint,epsf]{revtex}
\def\li{\mbox{\rm Li$_2$}}
\def\eps{\varepsilon}
\def\dot{\!\cdot\!}

\tighten
\begin{document}
\draft
\title{\hfill {\small DOE-ER-40757-112} \\
\hfill {\small UTEXAS-HEP-98-08} \\
\hfill  {\small MSUHEP-80622} \\
\hfill {\small MADPH-98-1061} \\
\hfill \\ 
$\bbox{\gamma\nu\to\gamma\gamma\nu}$ and crossed processes at energies below 
$\bbox{m_W}$} 

\author{Duane A. Dicus} 
\address{Center for Particle Physics and Department of Physics \\
 University of Texas, Austin, Texas 78712}

\author{Chung Kao}
\address{Department of Physics\\
 University of Wisconsin, Madison, Wisconsin 53706}

\author{Wayne W. Repko} 
\address{Department of Physics and Astronomy \\  
Michigan State University, East Lansing, Michigan 48824}
\date{\today}
\maketitle

\begin{abstract}
The cross sections for the processes $\gamma \nu\rightarrow \gamma
\gamma \nu$, $\gamma\gamma\rightarrow\gamma\nu\bar{\nu}$ and
$\nu\bar{\nu}\rightarrow\gamma\gamma\gamma$ are calculated for a range of 
center of mass energies from below $m_e$ to considerably above $m_e$, but much 
less than $m_W$. This enables us to treat the neutrino--electron coupling as a 
four--Fermi interaction and results in amplitudes which are electron box 
diagrams with three real photons and one virtual photon at their vertices. 
These calculations extend our previous low--energy effective interaction 
results to higher energies and enable us to determine where the effective 
theory is reliable.
\end{abstract}

\section{Introduction}

The $2\to 2$ processes $\gamma\nu\to\gamma\nu$, $\gamma\gamma\to\nu\bar{\nu}$
and $\nu\bar{\nu}\to\gamma\gamma$, which have potential astrophysical
applications, are known to be highly suppressed due to the vector--axial vector
nature of the weak interaction \cite{gell,yang,mjl,ls,cy,liu,dr93}. For massless
neutrinos, this suppression is such that the cross sections for the
$2\to 3$ processes $\gamma\nu\to\gamma\gamma\nu$, 
$\gamma\gamma\to\nu\bar{\nu}\gamma$ and $\nu\bar{\nu}\to\gamma\gamma\gamma$
exceed the $2\to 2$ cross sections for center of mass energies $2\omega$ between
1 keV and 1 MeV \cite{dr97}. While this range of energies is adequate for many
astrophysical applications, there are those, such as supernova dynamics,
where the $2\to 3$ cross sections at higher energies are needed \cite{hwt}.
The low--energy $2\to 3$ cross sections can be calculated from an effective
interaction \cite{dr97}, but these results become unreliable for energies on 
the order of $m_e$. Our purpose here is to compute these cross 
sections using the complete one--loop electron box amplitudes.

In the next Section, we discuss the relation between the low--energy 
description of $\gamma\nu\to\gamma\gamma\nu$ and its crossed channels based
on the effective interaction and the more exact numerical treatment of the full
amplitude. This is followed by a discussion of the numerical
results and their possible applications. The Appendix contains an expression for
the photon--photon scattering scalar four--point function in the case when one 
of the photons is virtual.

\section{$\bbox{2\to 3}$ processes and virtual photon--photon scattering}

The $2\to 3$ process $\gamma\nu\to\gamma\gamma\nu$ and its crossed channels can
be calculated using the diagrams illustrated in Fig.\,1. When the center of mass
energy is small compared to $m_W$, the $W$ and $Z$ propagators can be replaced
by $m_W^{-2}$ or $m_Z^{-2}$. In this case, the diagrams reduce to those for 
photon-photon scattering with one photon polarization vector replaced by the
neutrino current $\bar{u}(p_2)\gamma_{\mu}(1 + \gamma_5)u(p_1)$ or its crossed
channel counterparts. This reduction is illustrated in Fig.\,2 for the various 
channels.

For low--energy scattering, it is possible to describe all $2\to 3$ processes 
using the effective local interaction \cite{dr97}
\begin{equation} \label{l3gam}
{\cal L}_{\rm eff} = 4\frac{G_F\,a}{\sqrt{\displaystyle 2}}
\frac{\alpha^{3/2}}{\sqrt{\displaystyle 4\pi}}
\frac{1}{m_e^4}\left[\frac{5}{180}\left(N_{\mu\nu}F_{\mu\nu}
\right)\left(F_{\lambda\rho}F_{\lambda\rho}\right) - \frac{14}{180}N_{\mu\nu}
F_{\nu\lambda}F_{\lambda\rho}F_{\rho\mu}\right]\,,
\end{equation}
where $N_{\mu\nu}$ is
\begin{equation} 
N_{\mu\nu} = \partial_{\mu}\left(\bar{\psi}\gamma_{\nu}(1 + \gamma_5)\psi
\right) 
- \partial_{\nu}\left(\bar{\psi}\gamma_{\mu}(1 + \gamma_5)\psi\right)\,.
\end{equation}
Here, $a = \case{1}{2} + 2\sin^2\theta_W$ includes both the $W$ and $Z$
contributions.
The dimension 10 operator 
${\cal L}_{\rm eff}$ is closely related to the Euler--Heisenberg Lagrangian for 
photon-photon scattering \cite{e-h} and gives the leading term of an expansion
in $\omega/m_e$. The resulting low--energy cross section, 
\begin{equation} \label{inelas}
\sigma(2\to 3) = \frac{{\cal N}(2\to 3)}{637\;875}
\frac{G_F^{\,2}\,a^2\,\alpha^3}{\pi^4}\left(\frac{\omega}{m_e}\right)^8
\omega^2\,,
\end{equation}
exhibits a characteristic $\omega^{10}$ behavior for $\omega < m_e$. The 
numerical factors ${\cal N}(2\to 3)$ are
\begin{mathletters}
\begin{eqnarray}
{\cal N}(\gamma\nu\to\gamma\gamma\nu) & = & 1310\,, \\
{\cal N}(\gamma\gamma\to\nu\bar{\nu}\gamma) & = & 2144\,, \\
{\cal N}(\nu\bar{\nu}\to\gamma\gamma\gamma) & = & 952\,.
\end{eqnarray}
\end{mathletters}
When the center of mass energy exceeds $2m_e$, the box diagram develops an
imaginary part and it is no longer possible to obtain a reliable estimate of the
amplitude by expanding in inverse powers of $m_e$. As long as the center of mass
energy is small compared to $m_W$, it is still possible to calculate the
amplitude using the electron box diagrams of Fig.\,2 by treating the neutrino
current as a virtual photon. For the channels 
$\gamma\gamma\to\nu\bar{\nu}\gamma$ (b) and $\nu\bar{\nu}\to\gamma
\gamma\gamma$ (c), 
the virtual photon is timelike, while the virtual photon in the channel
$\gamma\nu\to\gamma\gamma\nu$ (a) is spacelike. In all cases, the diagrams 
can be expressed in terms of scalar two--point, three--point, and four--point 
functions and scalar products of external momenta, photon polarization vectors 
and the neutrino current \cite{pv}. The scalar functions, which are expressible
in terms of dilogarithms \cite{'thv}, were evaluated numerically using a 
FORTRAN code developed for one--loop corrections \cite{dk}.

\section{Discussion and conclusions}

The exact cross sections for all $2\to 3$ channels as a function of the center 
of mass energy of one of the initial particles are shown in Figs. 3--5, 
together with the low--energy approximation, Eq.\,(\ref{inelas}). In the
$\gamma\gamma\to\nu\bar{\nu}\gamma$ and $\nu\bar{\nu}\to\gamma\gamma\gamma$
cases, where the virtual photon is timelike, the low-energy result is valid up
to energies $\sim.3-.4m_e$. The deviation for a few higher energies is shown in
Table I. The ratio of the exact cross section to that given in Eq.\,
(\ref{inelas}) becomes unity at $\omega = 2.1m_e$ and $1.8m_e$ for the
$\nu\bar{\nu}$ and $\gamma\gamma$ channels, respectively. Below these
crossover points the ratio for these channels is always larger than unity.
Consequently, the effective interaction can be used up to these energies to set
lower bounds on physical effects.

For the remaining channel,
$\gamma\nu\to\gamma\gamma\nu$, where the virtual photon is spacelike, the 
agreement between the low--energy approximation and the complete calculation 
is good for energies as large as $m_e$. The reason for this difference is
related to the behavior of the amplitudes as the center of mass energy of the
each initial particle approaches $m_e$. For the timelike cases, which, in the
context of virtual ($\gamma^{\star}$) photon--photon scattering, correspond to 
$\gamma\gamma\to\gamma\gamma^{\star}$ and $\gamma^{\star}\to\gamma\gamma\gamma$,
$\omega\to m_e$ is precisely the threshold for $e\bar{e}$ production. This is 
the source of the cusp at $\omega = m_e$ in Figs.\,4 and 5, which is preceded 
by a departure from the low--energy $\omega^{10}$ behavior. In the spacelike 
case ($\gamma\gamma^{\star}\to\gamma\gamma$), the threshold for $e\bar{e}$ 
production requires
\begin{equation} \label{thresh}
E^{\prime}_{\nu}\leq E_{\nu} - \frac{m_e^2}{E_{\nu}}\,,
\end{equation}
where $E_{\nu}$ is the energy of the initial neutrino and $E^{\prime}_{\nu}$ is
the energy of the final neutrino. When $E_{\nu}\to m_e$, $E^{\prime}_{\nu}\to
0$, and there is essentially no phase space for this. Thus, the development of
an imaginary part for $\gamma\nu\to\gamma\gamma\nu$ occurs for energies
$E_{\nu}>m_e$ and the $\omega^{10}$ behavior persists to a higher value of
$\omega$. In this case, ratio of the exact cross section to Eq.\,(\ref{inelas}) 
is less than unity below $\omega = 1.03m_e$, grows to 1.71 at 
$\omega = 1.30m_e$ and again becomes unity at $\omega = 1.70m_e$.

The other feature of Figs.\,3--5 is the onset of an $\omega^2$
behavior once $\omega\gtrsim 50m_e$. This is expected since if $\omega >> m_e$, 
the only scale is $m_W$ or, equivalently, $G_F$ and hence the cross section 
should behave as $G_F^2\omega^2$. This behavior, together with the values
$\sigma_{\gamma\gamma\to\nu\bar{\nu}\gamma} = 5.68\times 10^{-49}$\,cm$^2$,
$\sigma_{\nu\bar{\nu}\to\gamma\gamma\gamma} = 4.13\times 10^{-49}$\,cm$^2$, and
$\sigma_{\gamma\nu\to\gamma\gamma\nu} = 1.74\times 10^{-48}$\,cm$^2$ at $\omega
= 100m_e$\,, allow accurate extrapolation to all higher energies much less than
$\omega = m_W$.

As in the low--energy case \cite{dr97}, the final photons in the 
$\gamma\nu\to\gamma\gamma\nu$ channel acquire circular polarization, which is
characteristic of a parity violating interaction. The magnitude of this effect
is illustrated in Fig.\,6, where the difference between the positive and
negative helicity cross sections for one of the final photons is plotted
relative to the total cross section. There is a reasonably large polarization,
$\sim$20--30\%, for center of mass energies $< 100\,m_e$.

In addition to the total cross sections, we have investigated the energy and
angular distribution of the final neutrino in the process
$\gamma\nu\to\gamma\gamma\nu$. The distribution $d\sigma/dE_{\nu}$ is shown in
Figs.\,7 and 8 for a variety of center of mass energies $\omega$ of the incident
photon. In all cases, the distribution of final neutrino energies rises until
the inequality Eq.\,(\ref{thresh}) is no longer satisfied, at which point it
rapidly drops. The evolution of the angular distribution $d\sigma/dz$, where $z$
is the cosine of the neutrino scattering angle, is shown in Figs.\,9 and 10.
At $\omega = m_e$, the distribution is peaked in the backward direction and this
gradually changes into a sharply forward peaked distribution at $\omega =
100 m_e$.

The processes described here could affect supernovas at several stages of the
explosion. Work toward including these reactions in a supernova code is in
progress. In addition, the scattering from infrared and optical backgrounds
given by these processes could attenuate travel of high energy photons and
neutrinos over cosmological distances. The cross sections are too small for
significant scattering from cosmic backgrounds.

Finally, it has been noted that the effective interaction, Eq.\,(\ref{l3gam}),
with one of the photons replaced by an external magnetic field could give
enhanced stellar cooling in stars with strong magnetic fields \cite{shai}. A
more exact calculation, along the lines of this paper, is under consideration.
\acknowledgements

We would like to thank V. Teplitz for helpful conversations. 
This research was supported in part by the
U. S. Department of Energy under Grants No. DE-FG02-95ER40896,
DE-FG013-93ER40757, in part by the National Science Foundation under Grant No.
PHY-93-07780 and in part by the University of Wisconsin Research Committee with
funds granted by the Wisconsin Alumni Research Foundation.

\appendix
\section*{The scalar function $\bbox{D_0}$ for
$\bbox{\gamma\gamma\to\gamma\gamma^{\star}}$}

The general expression for $D_0(1,2,3,4)$ with a common internal mass $m$,
massless external particles $k_1,k_2, k_3$, an external particle 
with $k_4^2\neq 0$ and all incoming momenta is \cite{gm}
\begin{eqnarray}
D_0(1,2,3,4) & = & \int_0^1\frac{dx}{(2k_1\dot k_2)(2k_2\dot k_3)x(1 - x) -
m^2(2k_1\dot k_3)}\left[\ln\left(1 + \frac{2k_1\dot k_2}{m^2}x(1 - x) - i\eps
\right)\right. \nonumber \\ [4pt]
&   &\hspace{11pt}\left.
 + \ln\left(1 + \frac{2k_2\dot k_3}{m^2}x(1 - x) - i\eps\right) -
 \ln\left(1 + \frac{k_4^2}{m^2}x(1 - x) - i\eps\right)\right]\,.
\end{eqnarray}
If we introduce the variables $PX4, PX5$ and $PX6$ as \cite{pv}
$$
PX4 = k_4^2\qquad PX5 = (k_1 + k_2)^2\qquad PX6 = (k_2 + k_3)^2\;,
$$ 
and define $PX7$ as $PX7 = PX4 - PX5 - PX6$, we can write
\begin{eqnarray}
D_0(1,2,3,4) & = & \int_0^1\frac{dx}{PX5\,PX6\,x(1 - x) - m^2 PX7}
\left[\ln\left(1 + \frac{PX5}{m^2}x(1 - x) - i\eps\right)\right.\nonumber 
\\ [4pt]
&   &\hspace{11pt}\left.+ \ln\left(1 + \frac{PX6}{m^2}x(1 - x) - i\eps\right) 
 -\ln\left(1 + \frac{PX4}{m^2}x(1 - x) - i\eps\right)\right]\,.
\end{eqnarray}
In this form, it is clear that, barring cancelations, $D_0(1,2,3,4)$ has an 
imaginary part whenever one or more of the $PXi$'s satisfies $PXi < -4m^2,\; i =
4,5,6$.

For the standard decomposition, we define the roots of the polynomials in the
logarithms as
\begin{eqnarray}
\beta_{\pm} - i\eps_5 & = & \frac{1}{2}\left(1 \pm \sqrt{1 + 4m^2/PX5}\right) -
i\frac{|PX5|}{PX5}\eps \,, \\ [4pt]
\gamma_{\pm} - i\eps_6& = & \frac{1}{2}\left(1 \pm \sqrt{1 + 4m^2/PX6}\right) -
i\frac{|PX6|}{PX6}\eps \,, \\ [4pt]
\delta_{\pm} - i\eps_4& = & \frac{1}{2}\left(1 \pm \sqrt{1 + 4m^2/PX4}\right) -
i\frac{|PX4|}{PX4}\eps \,,
\end{eqnarray}
and the roots of the denominator as
\begin{equation}
\lambda_{\pm} = \frac{1}{2}\left(1 \pm \sqrt{1 - 4m^2\,PX7/PX5\,PX6}\right)\,.
\end{equation}
In terms of these roots, we have
\begin{eqnarray} \label{d0}
D_0(1,2,3,4) & = & \frac{-1}{PX5\,PX6\,(\lambda_+ - \lambda_-)}\left\{
- \li\left(\frac{1 - \lambda_+}{\beta_+ - \lambda_+ - i\eps_5}\right)
+ \li\left(\frac{- \lambda_+}{\beta_+ - \lambda_+ - i\eps_5}\right)\right. 
\nonumber \\
&   &\left.- \li\left(\frac{1 - \lambda_+}{\beta_- - \lambda_+ + i\eps_5}\right)
+ \li\left(\frac{- \lambda_+}{\beta_- - \lambda_+ + i\eps_5}\right)
- \li\left(\frac{1 - \lambda_+}{\gamma_+ - \lambda_+ - i\eps_6}\right)\right. 
\nonumber \\ [4pt]
&   &\left.+ \li\left(\frac{- \lambda_+}{\gamma_+ - \lambda_+ - i\eps_6}\right)
- \li\left(\frac{1 - \lambda_+}{\gamma_- - \lambda_+ + i\eps_6}\right)
+ \li\left(\frac{- \lambda_+}{\gamma_- - \lambda_+ + i\eps_6}\right)\right. 
\nonumber \\ [4pt]
&   &\left.+\li\left(\frac{1 - \lambda_+}{\delta_+ - \lambda_+ - i\eps_4}\right)
-\li\left(\frac{- \lambda_+}{\delta_+ - \lambda_+ - i\eps_4}\right)
+\li\left(\frac{1 - \lambda_+}{\delta_- - \lambda_+ + i\eps_4}\right)\right. 
\nonumber \\ [4pt]
&   &\left.-\li\left(\frac{- \lambda_+}{\delta_- - \lambda_+ + i\eps_4}\right)
+ \ln\left(\frac{1 - \lambda_+}{-\lambda_+}\right)\left[
\mbox{\rule[-4pt]{0pt}{22pt}}\ln\left(-\frac{|PX5|}{PX5}
\frac{(PX4 - PX5)}{PX6} + i\eps_5\right)\right.\right.
\nonumber \\ [4pt]
&   &\left.\left.- i\pi\theta(PX5) + \ln\left(-\frac{|PX6|}{PX6}
\frac{(PX4 - PX6)}{PX5} + i\eps_6\right) - i\pi\theta(PX6)
\right.\right. \nonumber \\ [4pt]
&   &\left.\left.- \ln\left(-\frac{|PX4|}{PX4}\frac{(PX4 - PX6)}{PX5}
\frac{(PX4 - PX5)}{PX6} + i\eps_4\right) + i\pi\theta(PX4)
\mbox{\rule[-4pt]{0pt}{22pt}}\right]\right.\nonumber \\ [4pt]
&   &\left.\mbox{\rule[-4pt]{0pt}{22pt}} - \Bigl(\lambda_+\rightarrow 
\lambda_-\Bigr)\right\}\,,
\end{eqnarray}
where the dilogarithm, or Spence function $\li(z)$ is defined as
\begin{equation}
\li(z) = -\int_0^1\frac{dt}{t}\ln(1 - zt)\,.
\end{equation}
For numerical evaluation, $\li(z)$ can be expanded in powers of $-\ln(1 - z)$ 
\cite{'thv}. The contribution of the logarithims in the square bracket of Eq.\,
(\ref{d0}) is, at most, a phase.

\newpage

\begin{table}[h]
\begin{tabular}{cccc} 
$\omega/m_e$   & $\nu\bar{\nu}$ & $\gamma\gamma$ & $\gamma\nu$ \\ \hline
0.4      & 1.38   & 1.15  & 0.916 \\
0.5      & 1.68   & 1.27  & 0.883 \\
0.6      & 2.20   & 1.49  & 0.850 \\
0.7      & 3.17   & 1.79  & 0.823 \\
0.8      & 5.31   & 2.41  & 0.809 \\
0.9      & 11.9   & 3.95  & 0.815 \\
1.0      & 176.   & 23.3  & 0.877
\end{tabular}
\caption{The ratio of the exact cross section to that given by Eq.\,
(\ref{inelas}) for the various initial particles in the $2\to 3$ reactions is
shown.}
\end{table}

\begin{figure}[h]
\hspace{0.9in}
\epsfysize=1.8in \epsfbox{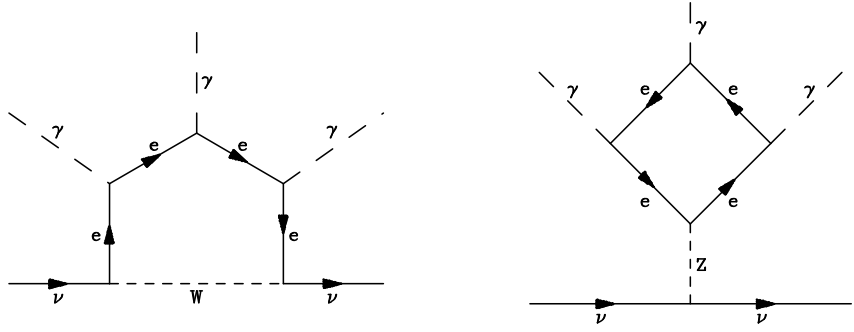}
\vspace*{8pt}
\caption{Typical diagrams for the process $\gamma\nu\protect\rightarrow\gamma
\gamma\nu$ arising from $W$ (left) and $Z$ (right) exchange are 
shown. The complete set is obtained by permuting the photons.}
\end{figure}

\begin{figure}[h]
\hspace{1.6in}
\epsfysize=2.2in \epsfbox{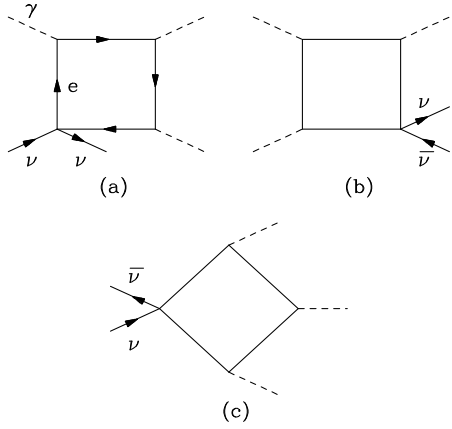}
\vspace*{8pt}
\caption{Typical box diagrams obtained in the limit of large $m_W$ are shown.}
\end{figure}

\begin{figure}[h]
\hspace{1.2in}
\epsfysize=2.1in \epsfbox{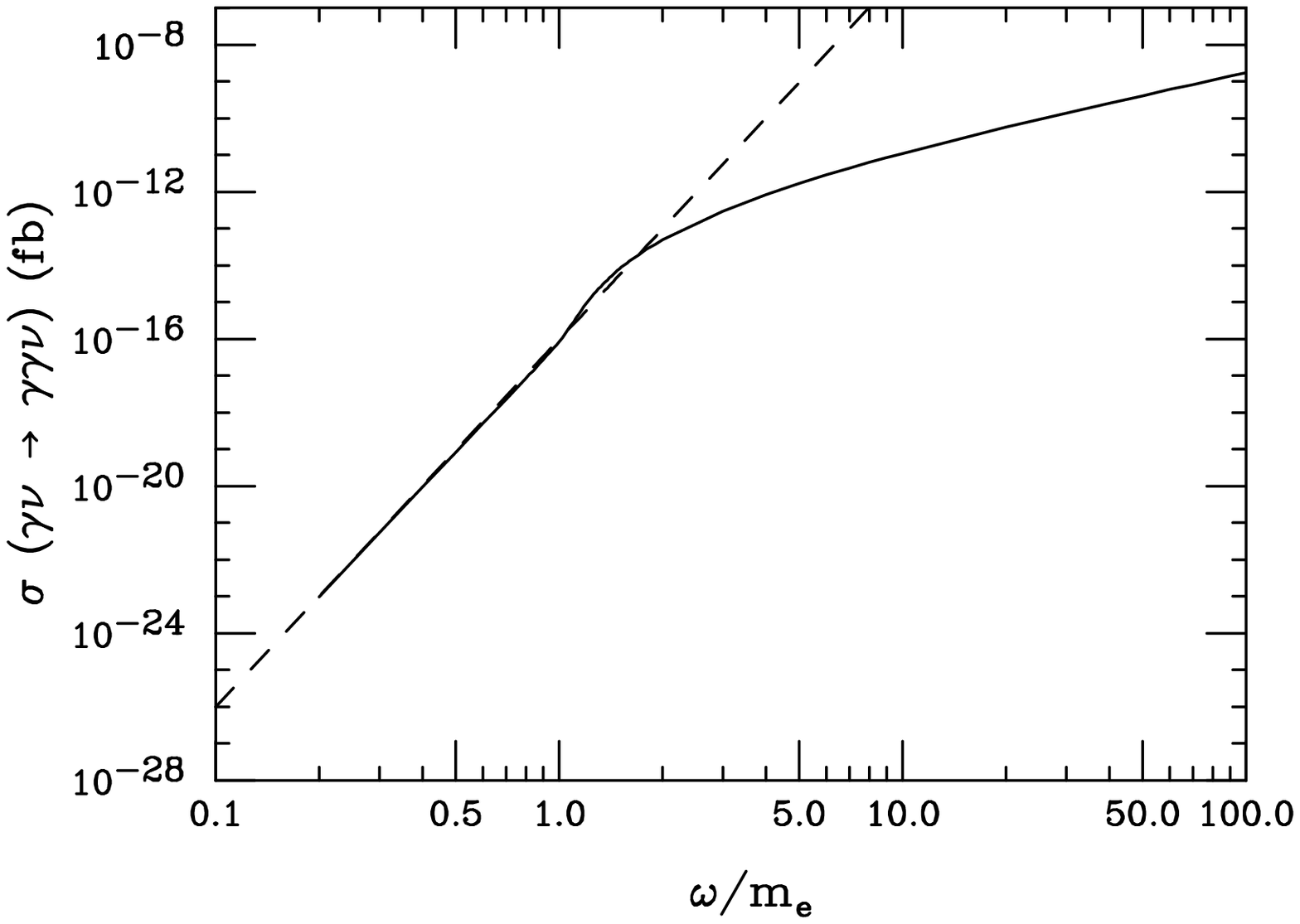}
\vspace*{8pt}
\caption{The cross section $\sigma(\gamma\nu\protect\to\gamma\gamma\nu)$ 
shown as the solid line. The dashed line is the low--energy cross section.}
\end{figure}

\newpage

\begin{figure}[h]
\hspace{1.2in}
\epsfysize=2.1in \epsfbox{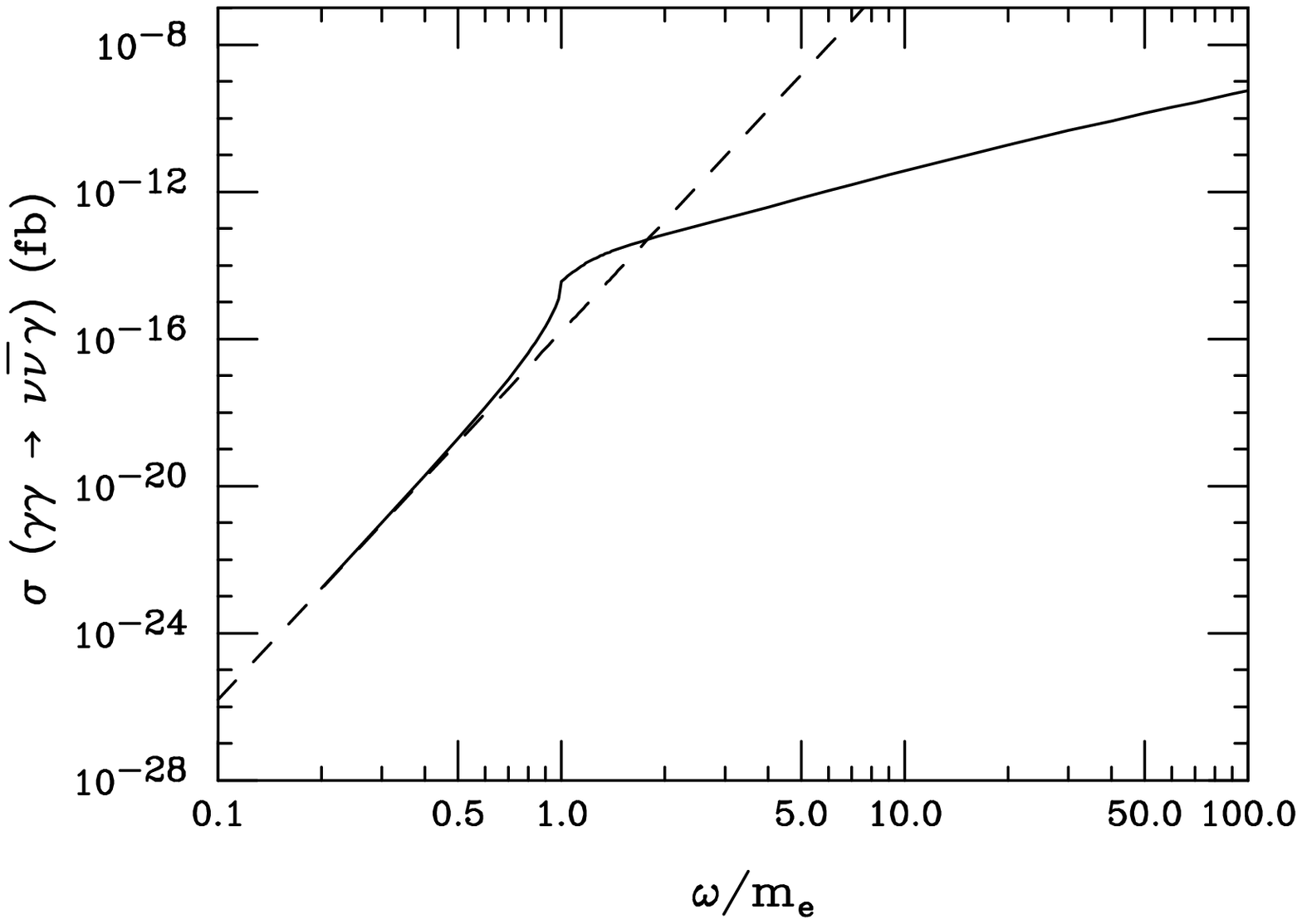}
\vspace*{8pt}
\caption{The cross section $\sigma(\gamma\gamma\protect\to\nu\bar{\nu}\gamma)$ 
shown as the solid line. The dashed line is the low--energy cross section.}
\end{figure}

\begin{figure}[h]
\hspace{1.2in}
\epsfysize=2.1in \epsfbox{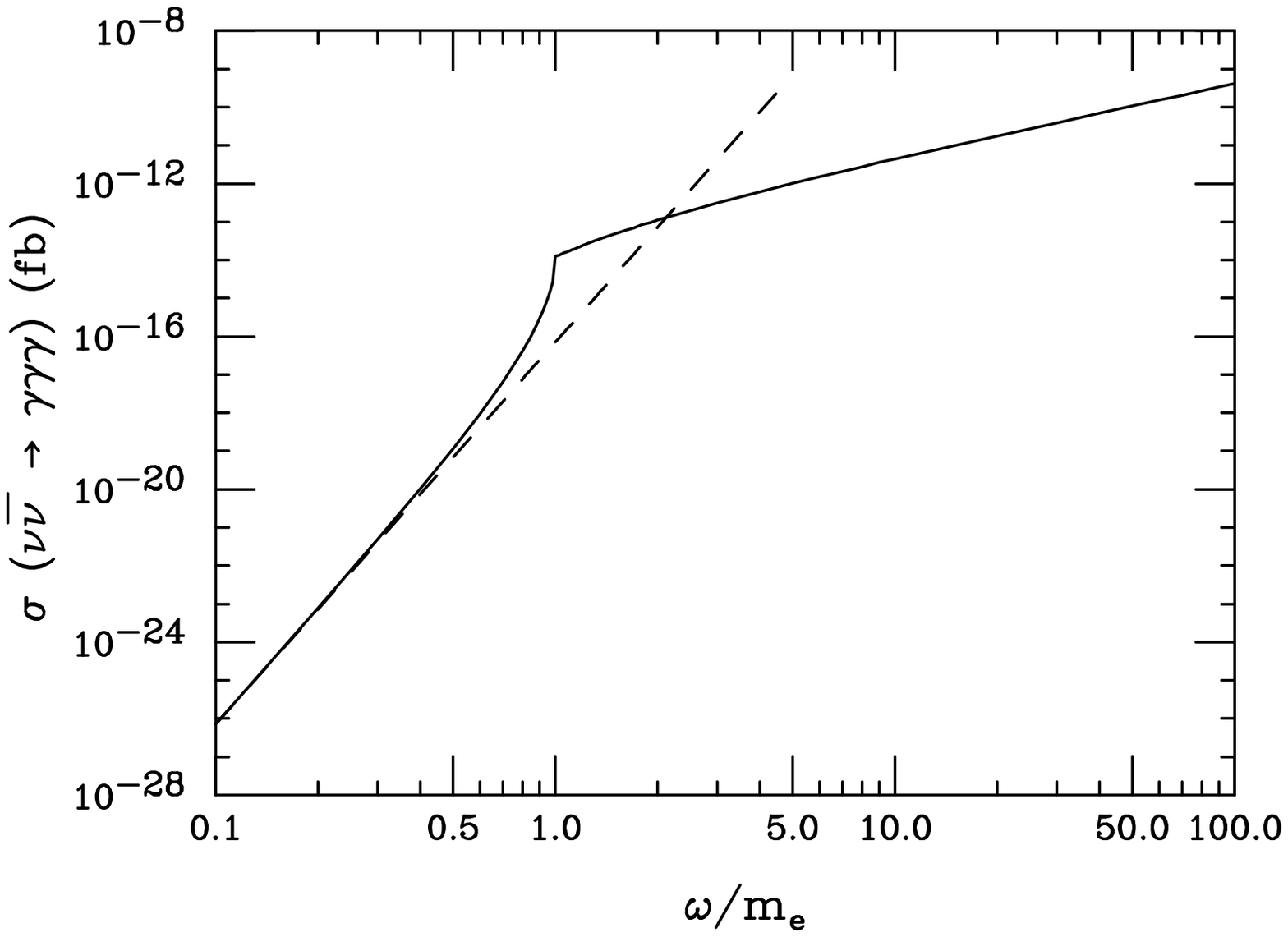}
\vspace*{8pt}
\caption{The cross section $\sigma(\nu\bar{\nu}\protect\to\gamma\gamma\gamma)$ 
shown as the solid line. The dashed line is the low--energy cross section.}
\end{figure}

\begin{figure}[h]
\hspace{1.2in}
\epsfysize=2.1in \epsfbox{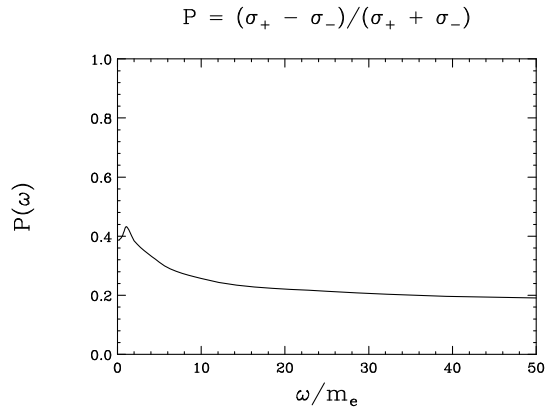}
\vspace*{8pt}
\caption{The circular polarization of one of the final photons in the process 
$\gamma\nu\protect\to\gamma\gamma\nu$ is shown.}
\end{figure}

\newpage

\begin{figure}[h]
\hspace{1.0in}
\epsfysize=2.1in \epsfbox{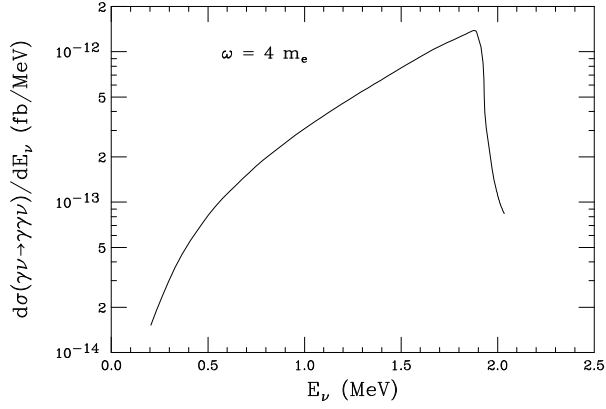}
\vspace*{8pt}
\caption{The distribution of final neutrino energies $d\sigma/dE_{\nu}$ is shown
for a photon center of mass energy $\omega = 4m_e$.}
\end{figure}

\begin{figure}[h]
\hspace{1.2in}
\epsfysize=2.1in \epsfbox{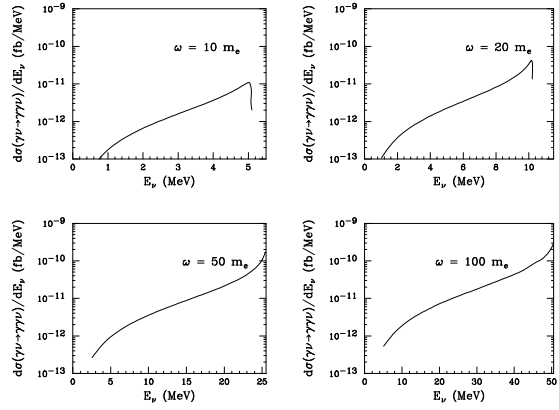}
\vspace*{8pt}
\caption{Same as Fig.\,7 for several values of photon center of mass energy
$\omega$.}
\end{figure}

\begin{figure}[h]
\hspace{0.9in}
\epsfysize=2.1in \epsfbox{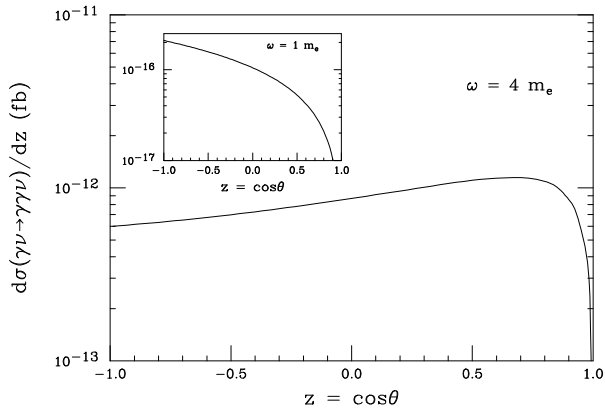}
\vspace*{8pt}
\caption{The angular distribution $d\sigma/dz$ is shown for $\omega = 4 m_e$.
The inset is the threshold ($\omega = m_e$) distribution, which is very nearly
$(1 - z)$.}
\end{figure}

\begin{figure}[h]
\hspace{1.2in}
\epsfysize=2.1in \epsfbox{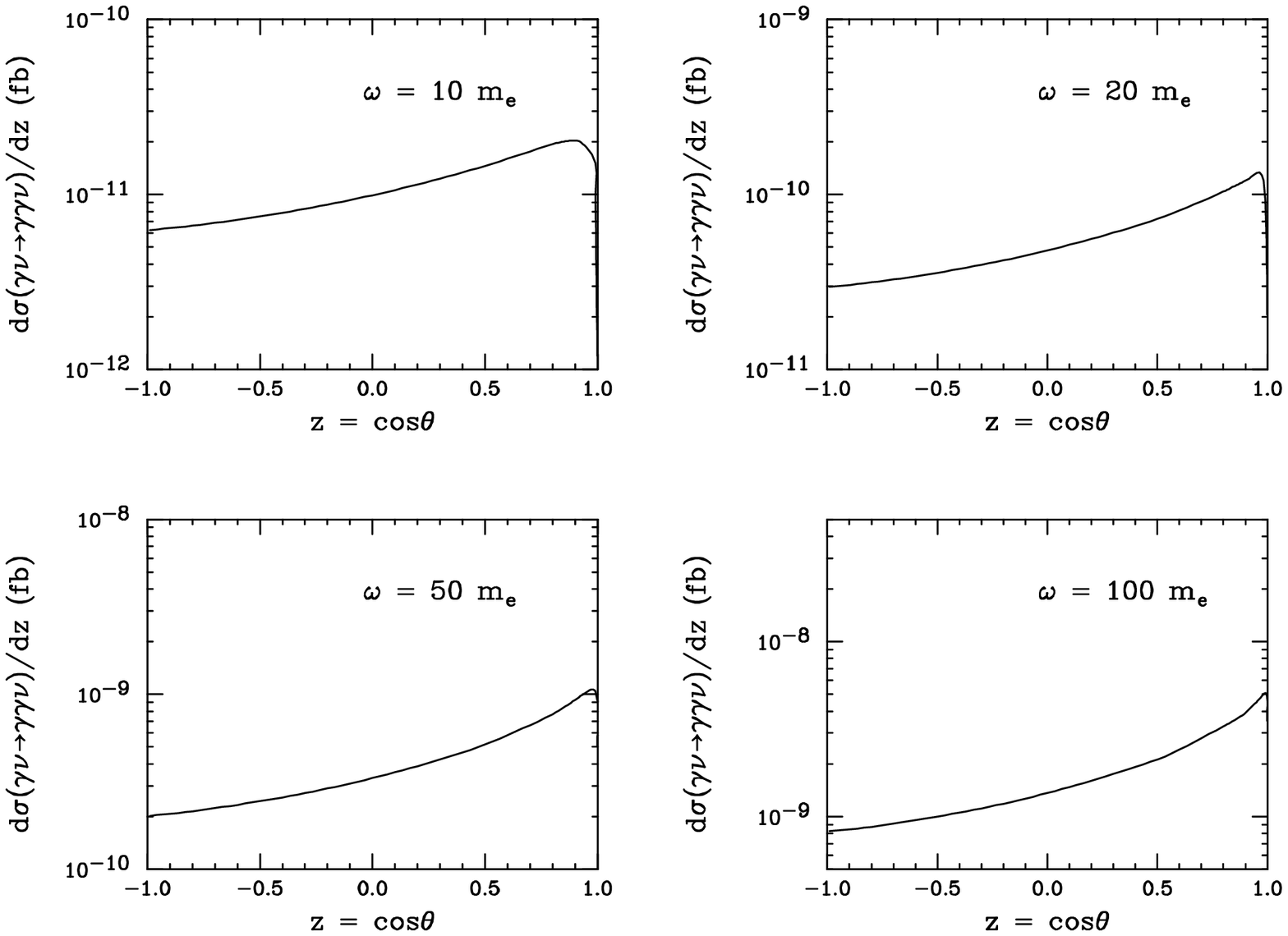}
\vspace*{8pt}
\caption{The angular distribution $d\sigma/dz$ is shown for several values of 
$\omega$.}
\end{figure}


\begin{thebibliography}{99}

\bibitem{gell} M. Gell-Mann, Phys. Rev. Lett. {\bf 6}, 70 (1961).
\bibitem{yang} C. N. Yang, Phys. Rev. {\bf 77}, 242 (1950); L. D. Landau,
Sov. Phys. Doklady {\bf 60}, 207 (1948).
\bibitem{mjl} M. J. Levine, Nuovo Cimento {\bf 48A}, 67 (1967).
\bibitem{ls} L. F. Landovitz and W. M. Schreiber, Nuovo Cimento {\bf 2A}, 359
(1971).
\bibitem{cy} V. K. Cung and M. Yoshimura, Nuovo Cimento {\bf 29A}, 557 (1975).
\bibitem{liu} J. Liu, Phys. Rev. D {\bf 44}, 2879 (1991).
\bibitem{dr93} D. A. Dicus and W. W. Repko, Phys. Rev. D {\bf 48}, 5106 (1993).
\bibitem{dr97} D. A. Dicus and W. W. Repko, Phys. Rev. Lett. {\bf 79}, 569
(1997). The channel $\gamma\gamma\to\nu\bar{\nu}\gamma$ is discussed in: N. Van
Hieu and E. P. Shabalin, Sov. Phys. JETP {\bf 17}, 681 (1963). 
\bibitem{hwt} M. Harris, J. Wang and V. Teplitz, astro-ph/9707113 (unpublished).
\bibitem{e-h} H. Euler, Ann. Phys. {\bf 26}, 398 (1936); W. Heisenberg and H.
Euler, Zeit. Phys. {\bf 98}, 714 (1936). The connection between these papers and
Eq.\,(\ref{l3gam}) can be seen by noting that the Hamiltonian is ${\cal H} =
-j\dot A - \ell\dot W - \ell^{\dagger}\dot W^{\dagger}$, where
$\ell_{\mu}$ is the lepton weak current. The the amplitude for pentagon 
diagram of Fig.\,1 involves the time ordered product $T\{\ell(x_1)\dot W(x_2)
\ell^{\dagger}(x_2)\dot W^{\dagger}(x_2)j(x_3)\dot A(x_3)j(x_4)\dot A(x_4)
j(x_5)\dot A(x_5)\}$ together with a combinatorial factor of 20. Contracting the
$W$ fields using the large mass limit of the $W$ propagator and Fierz
rearranging the weak currents gives 4 times the usual expression for the 
photon-photon scattering amplitude with one photon field replaced by the 
neutrino current. In addition, there is a term with one axial vector current,
which vanishes by Furry's theorem. Eq.\,(\ref{l3gam}) then follows. For
another discussion see: A. Abada, J. Matias and R. Pittau, hep-ph/9806383 
(1998).
\bibitem{pv} G. Passarino and M. Veltman, Nucl. Phys. {\bf B160}, 151 (1979).
\bibitem{'thv} G. 't Hooft and M. Veltman, Nucl. Phys. {\bf B153}, 365 (1979).
\bibitem{dk} C. Kao and D. A. Dicus, LOOP, a FORTRAN program for evaluating loop
integrals based on the results in Refs.\,\cite{pv} and \cite{'thv}.
\bibitem{shai} R. Shaisultanov, Phys. Rev. Lett. {\bf 80}, 1586 (1998).
\bibitem{gm} An expression for $D_0$ in the case when $k_4$ is timelike is
contained in E. W. N. Glover and A. G. Martin, Z. Phys. {\bf C60}, 175 (1993).
\end{thebibliography}
\end{document}